\documentclass[runningheads]{llncs}
\usepackage[T1]{fontenc}
\usepackage{graphicx}
\usepackage{multirow}
\usepackage{color}
\usepackage{threeparttable}
\usepackage{hyperref}

\begin{document}
%


\title{From First Draft to Final Insight: A Multi-Agent Approach for Feedback Generation}

\titlerunning{A Multi-Agent Approach for Feedback Generation}
%
 \author{Jie Cao\inst{1}\and
 Chloe Qianhui Zhao\inst{2}\and
 Xian Chen\inst{3}\and
 Shuman Wang\inst{4}\and
 Christian Schunn \inst{1}\and
 Kenneth R. Koedinger\inst{2}\and
 Jionghao Lin\inst{5,2 }\thanks{Corresponding author.}
 }
\authorrunning{J. Cao et al.}

\institute{
    University of Pittsburgh, Pittsburgh PA 15260, USA \\
    \email{{\{JIC319, schunn\}@pitt.edu}} \and
    Carnegie Mellon University, Pittsburgh PA 15213, USA \\
    \email{{\{cqzhao, koedinger, jionghao\}@cmu.edu}} \and
    The University of Manchester, Oxford Rd Manchester M139PL, UK \\
    \email{xian.chen-3@postgrad.manchester.ac.uk} \and
    Stanford University, Stanford CA 94305, USA \\
    \email{shuman@stanford.edu} \and
    The University of Hong Kong, Pokfulam Rd, Hong Kong, China\\
    \email{jionghao@hku.hk} 
}
\maketitle              

\begin{abstract}
Producing large volumes of high-quality, timely feedback poses significant challenges to instructors. To address this issue, automation technologies—particularly Large Language Models (LLMs)—show great potential. However, current LLM-based research still shows room for improvement in terms of feedback quality. Our study proposed a multi-agent approach performing ``generation, evaluation, and regeneration'' (G-E-RG) to further enhance feedback quality. In the first-generation phase, six methods were adopted, combining three feedback theoretical frameworks and two prompt methods: zero-shot and retrieval-augmented generation with chain-of-thought (RAG\_CoT). The results indicated that, compared to first-round feedback, G-E-RG significantly improved final feedback across six methods for most dimensions. Specifically:(1) Evaluation accuracy for six methods increased by $3.36\%$ to $12.98\%$ ($p<0.001$); (2) The proportion of feedback containing four effective components rose from an average of $27.72\%$ to an average of $98.49\%$ among six methods, sub-dimensions of providing critiques, highlighting strengths, encouraging agency, and cultivating dialogue also showed great enhancement ($p<0.001$); (3) There was a significant improvement in most of the feature values ($p<0.001$), although some sub-dimensions (e.g., strengthening the teacher-student relationship) still require further enhancement; (4) The simplicity of feedback was effectively enhanced ($p<0.001$) for three methods.

\end{abstract}
\section{Introduction}
Feedback is widely recognized as a pivotal element of student learning \cite{carless2024towards}. However, although extensive research has shown the positive influence of feedback on learning outcomes, the extent of this influence varies \cite{wisniewski2020power,hattie2007power}: some studies report that feedback significantly improves performance \cite{drouvelis2022feedback}, while others find less measurable difference \cite{goller2024good}. These disparities are due to multiple factors, including the timing of the feedback, its delivery format, and, most importantly, the quality of its content \cite{drouvelis2022feedback,kluger1996effects}. High-quality feedback is grounded in sound educational theory rather than merely providing correct answers. However, in practice, offering such personalized and theoretically robust feedback to large student cohorts is challenging due to the substantial effort required of instructors \cite{cavalcanti2021automatic,liang2024towards}. This challenge has driven interest in automated feedback generation.  Numerous studies have examined LLM-based feedback generation (e.g. \cite{dai2024assessing,guo2024resist}), and some have even explored its effects on student perceptions and learning outcomes in real educational settings (e.g. \cite{gabbay2024combining,nguyen2024comparing}). Despite these advances, using LLMs to generate high-quality feedback faces several limitations. First, LLMs trained in the Internet corpora can produce hallucinations in specialized domains \cite{Huang-2025}. Second, they tend to generate lengthy or verbose responses \cite{saito2023verbosity}, which can reduce clarity and efficiency. Third, the feedback they produce often lacks key pedagogical elements -particularly those that support self-regulation and independent learning \cite{dai2024assessing} - that are increasingly important as education shifts from a teacher-directed model to a learner-focused model \cite{dunbar2022shifting}. To address the first two issues, various prompt engineering techniques (ranging from zero-shot and few-shot approaches to advanced methods such as CoT and RAG \cite{sahoo2024systematic}) can be applied. Meanwhile, integrating robust educational theories (for example, those that emphasize learner-centered feedback framework \cite{ryan2021designing}) can help overcome the third limitation. More importantly, employing a multi-agent, multi-round process may help mitigate the inherent instability of single-round feedback generation by enabling LLMs to collaborate, thus further addressing the aforementioned limitations \cite{wu2023autogen}. This study, therefore, aims to explore a multi-agent system that employs a ``generation, evaluation, and regeneration'' (G-E-RG) process, where initial feedback is generated by different combinations of prompt engineering techniques and feedback frameworks, then is automatically evaluated and refined. The research questions are as follows.
\begin{itemize}
\item \textit{RQ1: Which combination of prompt techniques and feedback framework generates the initial highest-quality feedback?}
\item \textit{RQ2: How accurately can LLMs evaluate the generated feedback?}
\item \textit{RQ3: To what extent does the ``generation, evaluation and regeneration'' method improve the quality of LLM-generated feedback?}
\end{itemize}
\section{Related Work}
\subsubsection{Conceptualizations of Feedback}
Feedback helps learners identify their current performance and close gaps between actual and desired outcomes \cite{hattie2007power,ramaprasad1983definition}. As a critical educational component, feedback significantly influences student learning \cite{hattie2007power,poulos2008effectiveness}. A meta-analysis revealed moderate but varied effects, indicating that feedback effectiveness depends largely on its quality and design \cite{wisniewski2020power}. Effective feedback aligns with established educational theories and frameworks, categorized into knowledge transmission-centered models—focusing on transferring information from teacher to learner—and learner-centered models, emphasizing learner agency and dialogue-based interactions \cite{ryan2021designing,liang2024towards,hattie2007power}. High-quality feedback should be accurate, clear, concise, and reflective of learner-centered principles, transforming traditional teacher-driven comments into interactive dialogues that foster active student engagement \cite{da2024analysis,dai2024assessing,ryan2021designing}.

\subsubsection{Feedback Generation using LLMs}
Recent advances in artificial intelligence have shifted feedback generation from manual generated and rule-based methods to LLM-facilitated approaches, demonstrating effectiveness across contexts from K-12 to higher education, including computer science \cite{nguyen2024comparing,fung2024automatic} and language learning \cite{teng2024chatgpt,guo2024resist}. For instance, ChatGPT-enhanced feedback increased motivation, self-efficacy, and engagement in English language learners \cite{teng2024chatgpt}, and improved clarity and personalization in K-12 data science education \cite{fung2024automatic}. Nevertheless, challenges such as accuracy persist, with LLMs frequently generating incorrect or incomplete feedback \cite{gabbay2024combining,nguyen2024comparing,chen2024multi}. Although generally readable \cite{dai2024assessing}, AI-generated feedback can be excessively lengthy \cite{fokides2024comparing} and often falls short in quality, lacking effective elements, such as feeding-up information and self-regulation guidance, compared to human-generated feedback \cite{dai2024assessing}. Moreover, it tends toward direct answers rather than promoting interactive dialogue-based learning experiences. To address these limitations, advanced prompting techniques like retrieval-augmented generation (RAG) and Chain-of-Thought (CoT) reasoning can enhance precision and analytical depth in feedback \cite{lewis2020retrieval,wei2022chain}. In addition, integrating theoretical frameworks with RAG\_CoT could further improve feedback quality. Moreover, multi-agent mechanisms, effective in instructional contexts \cite{yu2024mooc}, code generation \cite{huang2023agentcoder}, further refine feedback generation quality \cite{wu2023autogen}.

\section{Methods}
\subsection{Course Context and Dataset}
Data was collected from a graduate course entitled \textit{E-Learning Design Principles and Methods}, offered at Carnegie Mellon University in the Fall of $2024$.  
 This course contains specialized educational knowledge that LLMs may not fully capture due to restricted access. Our dataset consists of $63$ multiple-choice questions (with $152$ anonymous responses) and one open-ended question (with $56$ anonymous responses) from the quiz of this course, totaling $208$ responses.

\subsection{Educational Framework of Feedback}
Our study adopted two frameworks to help LLMs generate feedback (the details are shown in \href{https://doi.org/10.5281/zenodo.14890689}{the Digital Appendix}. The first was the knowledge transmission-focused feedback framework from \cite{hattie2007power}, containing four levels: task, process, self-regulation, and self.  The second was the learner-centered feedback framework from \cite{ryan2021designing}, containing two dimensions: components and features. In addition, we adopted the non-feedback framework as our baseline framework.

\subsection{New Feedback Generation Method: ``G-E-RG''}
\begin{figure}
\includegraphics[width=\textwidth, trim=0cm 2.9cm 0cm 2.5cm, clip]{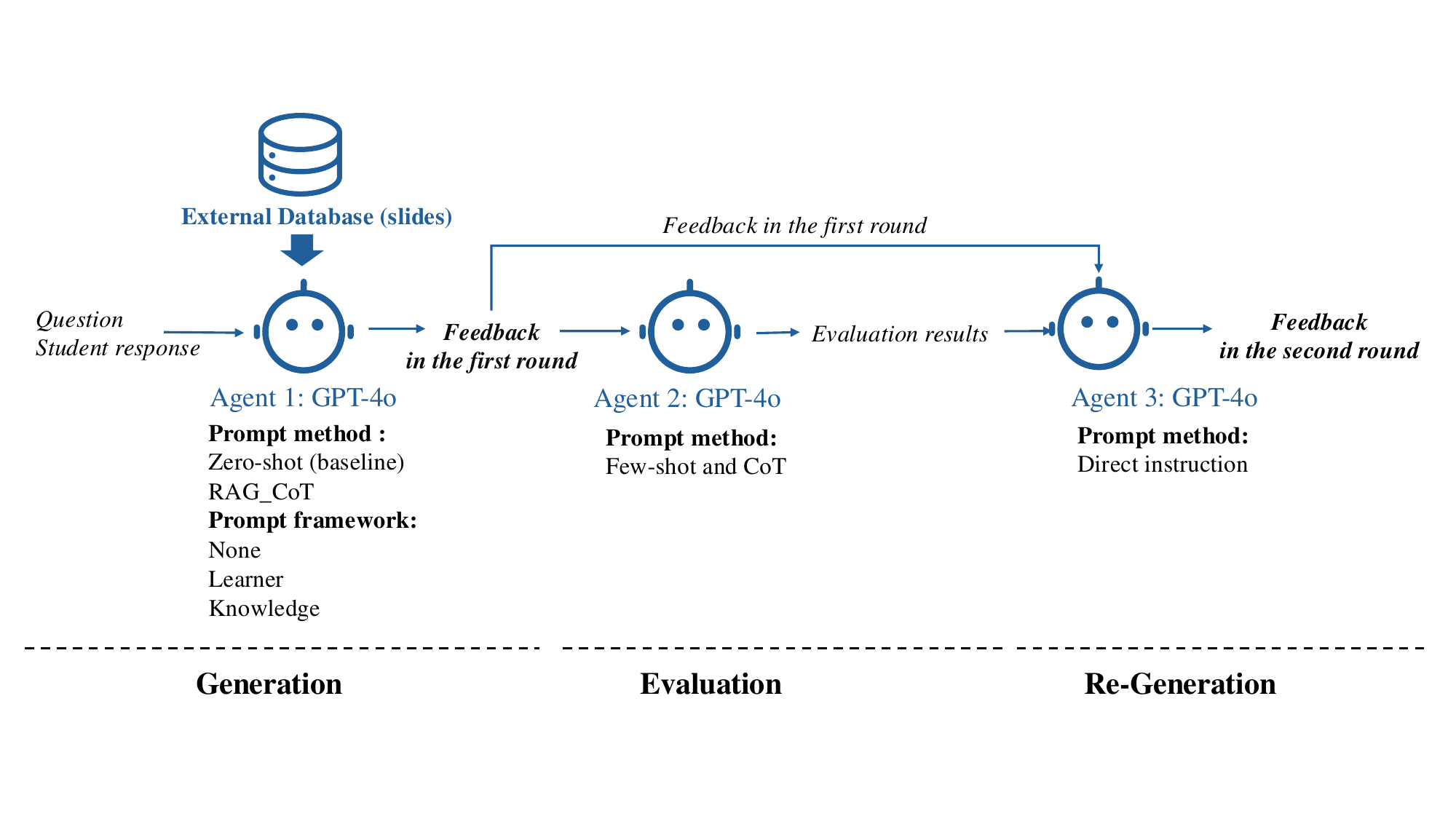}
\caption{The ``G-E-RG'' framework for feedback generation} \label{Fig1}
\end{figure}

Our study proposed a new feedback generation method named ``G-E-RG'' that used three agents to generate high-quality feedback for students' responses. The overall framework is shown in Fig.~\ref{Fig1}.
\subsubsection{Feedback Generation in the First Round} Our study used a\textbf{ }GPT-4o model to generate feedback in the first round. We put the quiz questions and corresponding students' responses from the course into the system along with prompts, which were combined with different prompt strategies: Zero-shot (as the baseline) and RAG\_CoT and different feedback frameworks: no framework, learner framework, knowledge framework. Six prompt example is shown in the \href{https://doi.org/10.5281/zenodo.14890689}{Digital Appendix}. Since there are six different prompt combinations, a total of $1248 $($6 \times 208$) pieces of feedback were generated.

\subsubsection{Feedback Rubric and Evaluation }  
\paragraph{The evaluation rubric of feedback} As shown in Table \ref{tab1}, we used two reliability indicators: evaluation accuracy and retrieved slide accuracy. These indicators assess whether the feedback accurately judges the student’s response and whether it accurately retrieves the relevant slides, respectively. Next, we adopted four components and five features to evaluate the effectiveness of feedback, following the learner-centered feedback framework proposed by \cite{ryan2021designing}, as it meets the requirements of learner-centered learning \cite{dai2024assessing,ryan2021designing}. Finally, we incorporated simplicity, measured by the word count of generated feedback. In combination with other measures, feedback with fewer words means more concise. The detailed rubric is presented in the \href{https://doi.org/10.5281/zenodo.14890689}{Digital Appendix}. 

\begin{table}
\scriptsize
\caption{The evaluation rubric of feedback quality.}\label{tab1}
\resizebox{1\textwidth}{!}{%
\renewcommand{\arraystretch}{1.5}
\begin{tabular}{l|p{8.6cm}|c}
\hline
\textbf{Dimension} & \textbf{Sub-Indicator} & \textbf{Value} \\
\hline
\multirow{2}{*}{Reliability}
 & {\itshape Evaluation accuracy} & True/False \\
 & {\itshape Retrieved slides accuracy} & \\
\hline
\multirow{4}{*}{Components}
 & {\itshape \textbf{C1-critiques}: Comments that provide critiques about performance}
 & \multirow{4}{*}{Yes/No} \\
 & {\itshape \textbf{C2-strengths}: Comments that highlight strengths of performance} & \\
 & {\itshape \textbf{C3-actionable}: Comments that provide actionable information for future performance} & \\
 & {\itshape \textbf{C4-agency}: Comments that encourage learner agency} & \\
\hline
\multirow{5}{*}{Features}
 & {\itshape \textbf{F1-positive}: Comments encourage positive learner affect}
 & \multirow{5}{*}{
  \begin{tabular}[c]{@{}l@{}}
    1 (No/Minimal)\\
    2 (Moderate)\\
    3 (Strong)
  \end{tabular}
}\\
 & {\itshape \textbf{F2-usable}: Comments are usable for learner} & \\
 & {\itshape \textbf{F3-relationship}: Comments strengthen teacher and learner relationships} & \\
 & {\itshape \textbf{F4-dialogue}: Comments invite dialogue about feedback} & \\
 & {\itshape \textbf{F5-independence}: Comments promote learner independence} & \\
\hline
Simplicity 
 & {\itshape The word count of the generated feedback}& 0-N \\
\hline
\end{tabular}
}
\end{table}

\paragraph{Human evaluation}  Two research assistants completed four iterative training cycles on the rubric until achieving high inter-rater agreement. They then double-coded 10\% of the feedback samples, obtaining Cohen’s Kappa values ranging from $0.751***$ to $0.962***$ across three dimensions: reliability, components and features. Afterward, each assistant independently coded a total of $1248$ feedback, serving as an evaluation result for the first round of feedback. Note that simplicity was calculated automatically by Python, requiring no human or LLMs-based evaluation.

\paragraph{Automatic evaluation based on LLMs} Parallel to the manual evaluation, we employed a GPT-4o model to automatically evaluate the feedback, which follows the same rubric used in the manual process. This phase used the few-shot and CoT as the prompt strategy.

\subsubsection{Feedback Re-Generation in the Second Round} 
After evaluating the feedback generated in the first round, this study decoded the evaluation results into suggestions, then put suggestions, questions, students' responses, and first-round feedback into another GPT-4o model to re-generate feedback. This study adopted the direct instruction prompt during the Re-Generation phase: ``\textit{Based on the question, student's response, previous feedback, and suggestion to modify feedback, generate the new feedback}''.  Note: Since the retrieved slides are determined based on similarity calculations, they are fixed. Therefore, the second round of generated feedback cannot improve the accuracy of retrieved slides and will still be based on the same slides for feedback generation. Apart from this, all other sub-indicators will be modified based on suggestions.

Finally, after feedback re-generation, to assess the quality of feedback in the second round, this study employed the GPT-4o model to automatically evaluate feedback, and let research assistants check the results. 

\section{Results}
\subsection{Comparison of quality of feedback generated in the first round among different methods}
This study compared the initial quality of feedback generated in the first round by six different methods across four dimensions: reliability, components, features, and simplicity. In terms of \textbf{reliability}, the RAG\_CoT condition outperforms the Baseline condition, with RAG\_CoT\_knowledge reaching the highest evaluation accuracy ($93.27\%$). The highest accuracy of retrieved slides in the RAG\_CoT condition is $94.23\%$, as shown in Table \ref{tab2}. 
\begin{table}
\centering
\scriptsize
\caption{The reliability of feedback generated by different methods}\label{tab2}
\renewcommand{\arraystretch}{1.5}
\begin{tabular}{l|l|c|c}
\hline
\textbf{Condition} & \textbf{Methods} &  \textbf{Accurate evaluations} & \textbf{Accurately retrieved slides}\\
\hline
Baseline   & Baseline\_none      & 185 (88.94\%)\textbf{ }& /\\
           & Baseline\_learner   & 171 (82.21\%) & /\\
           & Baseline\_knowledge & 174 (83.65\%) & /\\ 
RAG\_CoT   & RAG\_CoT\_none      & 186 (89.42\%) & \textbf{196 (94.23\%)}\\
           & RAG\_CoT\_learner   & 189 (90.87\%) & \textbf{196 (94.23\%)}\\
           & RAG\_CoT\_knowledge & \textbf{194 (93.27\%) }& 195 (93.75\%)\\
\hline
\end{tabular}
\end{table}

In terms of \textbf{components}, Table \ref{tab3} shows that both the Baseline and RAG\_CoT conditions using the learner framework yield higher coverage in highlighting strengths (C2, $96.15\%,93.30\%$), providing actionable suggestions (C3, $100\%,100\%$), and encouraging agency (C4, $100\%,97.11\%$), but low coverage in providing critiques (C1, $22.11\%,72.60\%$). In this component(C1-critiques), the knowledge condition performs better( with $85.58\%$ and $82.68\%$, respectively). Overall, the RAG\_CoT\_learner method generates more comprehensive feedback, with $62.98\%$ of its feedback including all four components, but there is still a big room for improvement, especially for other methods.
\begin{table}
\centering
\caption{The number and percentage of each component contained in the feedback generated by different methods}\label{tab3}
\resizebox{1\textwidth}{!}{%
\scriptsize
\renewcommand{\arraystretch}{1.5}
\begin{tabular}{l|c|c|c|c|c}
\hline
\textbf{Methods} &  \textbf{C1-critiques} & \textbf{C2-strengths}& \textbf{C3-actionable}& \textbf{C4-agency} & \textbf{All Components}\\
\hline
Baseline\_none &  48(23.08\%) &  131(62.98\%) & 195(93.75\%) &137(65.87\%) &22(10.58\%)\\
RAG\_CoT\_none&  56(26.92\%) &  111(53.36\%) & 166(79.80\%) & 138(66.35\%) &3(1.44\%)\\
Baseline\_learner &  46(22.11\%) &  200(96.15\%) & 208(100.00\%) & 208(100.00\%)& 43(20.67\%)\\
RAG\_CoT\_learner  &  151(72.60\%) &  194(93.30\%) & 208(100.00\%) & 202(97.11\%)& \textbf{131(62.98\%)}\\
Baseline\_knowledge &  178(85.58\%) &  87(41.83\%) & 202(97.12\%)& 144(69.23\%) & 66(31.73\%)\\
RAG\_CoT\_knowledge &  172(82.69\%) &  113(54.33\%) & 208(100.00\%) &182(87.50\%)&81(38.94\%)\\
\hline
\end{tabular}
}
\end{table}

\begin{figure}
\includegraphics[width=\textwidth, trim=0cm 0.8cm 0cm 0cm, clip]{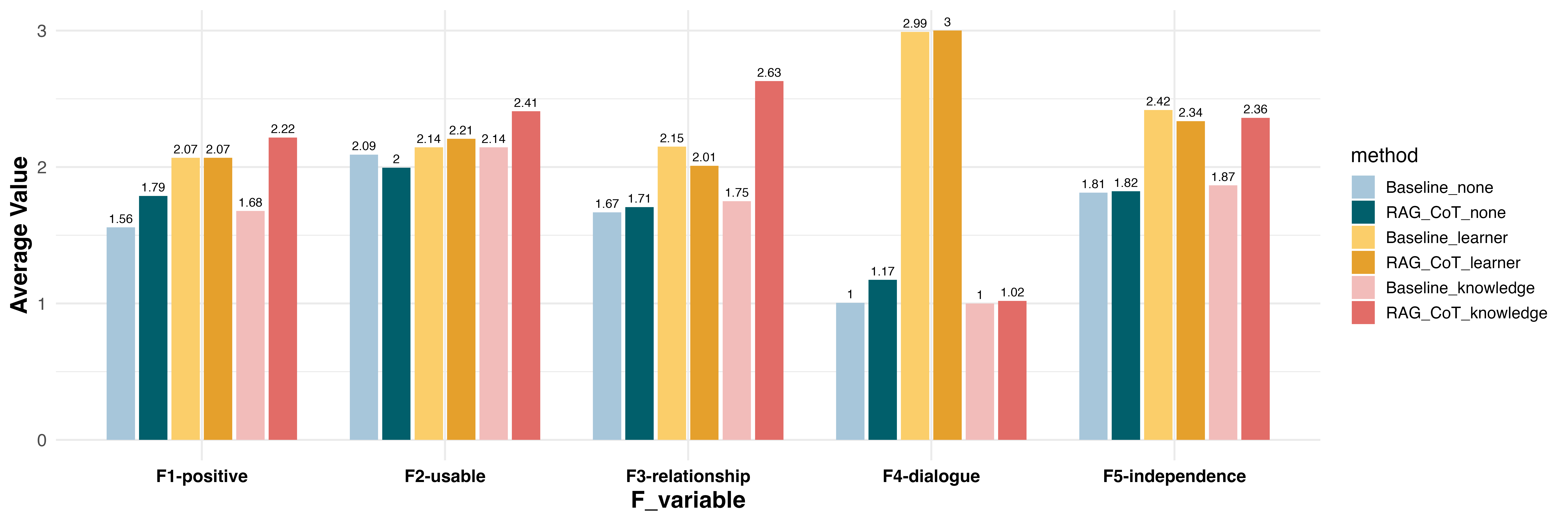}
\caption{The average value of five features of feedback generated by different methods} \label{fig2}
\end{figure}

Regarding \textbf{features}, Fig. \ref{fig2}  indicates that RAG\_CoT\_knowledge consistently scores highest in F1-positive ($2.22$), F2-usable ($2.41$), and F3-relationship ($2.63$). However, the learner framework (combined with Baseline and RAG\_CoT) outperforms the knowledge framework in F4-dialogue ($2.99$ and$ 3$, respectively) and F5-independence ($2.42$ and $2.34$, respectively). Overall, the RAG\_CoT\_learner and Baseline\_learner  method emerge as the more balanced method for generating feedback that keeps effective features, but there's also a need to strengthen the other four features, in addition to the F4-dialog.
\begin{figure}
\includegraphics[width=\textwidth, trim=0cm 0cm 0cm 1cm, clip]{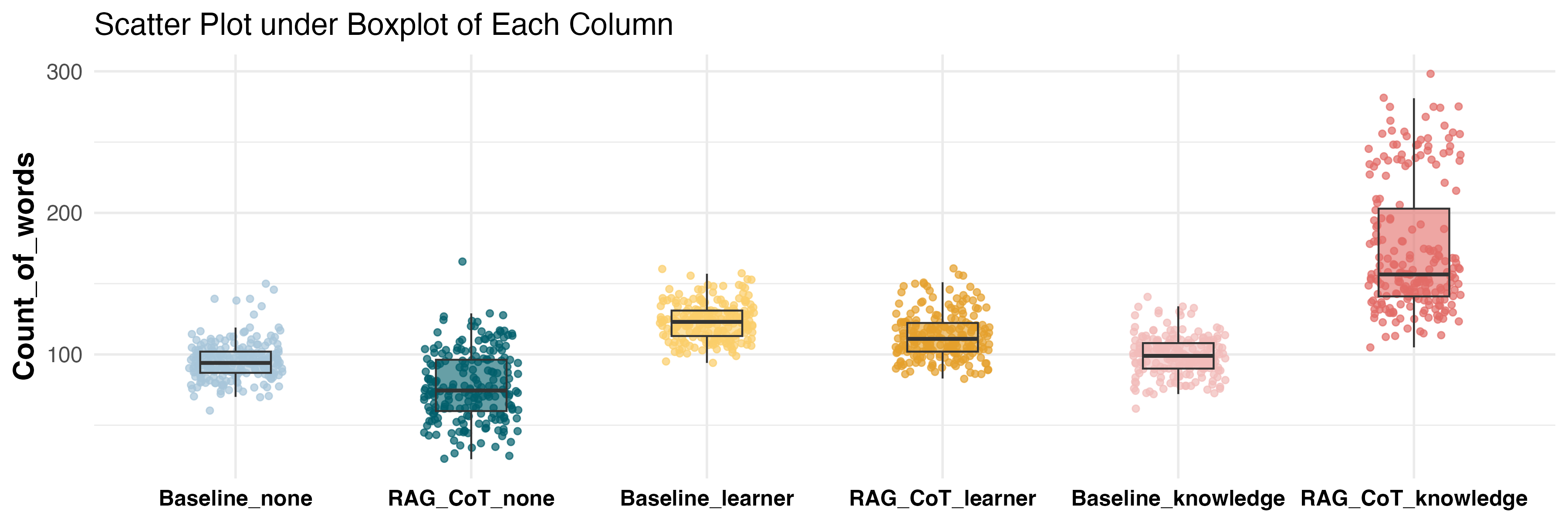}
\caption{The word count (simplicity) of feedback generated by different methods} \label{fig3}
\end{figure}
Regarding \textbf{simplicity}, as shown in Fig. \ref{fig3}, feedback generated without any framework was shorter—averaging $95.51$ words for baseline\_none and $78.01$ words for RAG\_CoT\_none. whereas incorporating a framework (learner or knowledge) led to longer responses; notably, RAG\_CoT\_knowledge produced the longest feedback, averaging $173.88$ words (SD=$46$),  necessitating further condensation in subsequent rounds of feedback generation. 
Our study found that the RAG\_CoT\_learner method produced the highest-quality feedback in the first round, but it still requires further improvement—let alone the first-round feedback generated by other methods.

%
%

\begin{table}[b!]
\centering
\caption{The precision, recall, F1 and accuracy for measuring the agreement between LLM and human coders}
\scriptsize
\label{tab4}
\renewcommand{\arraystretch}{1.5}
\begin{tabular}{l|l|c|c|c|c}
\hline
\textbf{Dimension} & \textbf{Sub-indicator} & \textbf{Accuracy} & \textbf{Precision} & \textbf{Recall} & \textbf{F1 Score} \\
\hline
\multirow{2}{*}{Reliability} & Evaluation accuracy  & 0.92 & 0.93 & 0.94 & 0.93 \\
                             & Retrieved slides accuracy & 0.86 & 0.87 & 0.86 & 0.84 \\
\hline
\multirow{4}{*}{Components} & C1-critiques& 0.86 & 0.87 & 0.86 & 0.86 \\
                             & C2-strengths& \textbf{0.97} & \textbf{0.97} &\textbf{ 0.97} & \textbf{0.97} \\
                             & C3-actionable& \textbf{0.98} & \textbf{0.99} & \textbf{0.98 }& \textbf{0.98} \\
                             & C4-agency& 0.89 & 0.89 & 0.89 & 0.88 \\
\hline
\multirow{5}{*}{Features}    & F1-positive& 0.82 & 0.84 & 0.82 & 0.79 \\
                             & F2-usable& 0.75 & 0.75 & 0.75 & 0.75 \\
                             & F3-relationship& 0.85 & 0.87 & 0.85 & 0.85 \\
                             & F4-dialogue& \textbf{0.99} & \textbf{0.99} & \textbf{0.99} & \textbf{0.99 }\\
                             & F5-independence& 0.71 & 0.76 & 0.71 & 0.73 \\
\hline

\end{tabular}
\end{table}

\subsection{Accuracy of feedback evaluation using LLMs}
Table \ref{tab4} presents the accuracy, precision, recall, and F1 score for assessing feedback quality across three dimensions (reliability, components and features) using LLMs.
For the \textbf{reliability dimension}, it performs medium to good, with F1 scores of $0.93$ and $0.84$, respectively.
Among the \textbf{components dimensions}, LLMs achieved the highest performance on C2-strengths (accuracy: $0.97$, F1 score: $0.97$) and C3-actionable (accuracy:$ 0.98$, F1 score: $0.98$). 
In terms of \textbf{features dimensions}, F4-dialogue demonstrated the highest accuracy ($0.99$). In addition, F1-positive and F3-Relationships performed moderately well, with F1 scores of $0.79$ and $0.85$, respectively.  F2-usable and F5-independence had the lowest F1 scores ($0.75$ and $0.73$), suggesting that the model struggles with reliably identifying these features. Overall, the LLMs demonstrated robust performance in assessing feedback in terms of  C2-strengths, C3-actionable and F4-dialogue, there is a shortcoming in terms of assessing other sub-indicators, especially F2-usable and F5-independence.

\subsection{The enhancement of feedback quality after the ``G-E-RG''}
This study compared the quality of first-round feedback with that of the regenerated second-round feedback across three dimensions to evaluate the effectiveness of the ``G-E-RG'' method. Regarding the \textbf{reliability} dimension (see Table \ref{tab5}), the regenerated feedback improved the evaluation accuracy (how much feedback accurately evaluates students' responses) for all methods. The most substantial increases were observed in the Baseline\_learner ($12.98\%$) and Baseline\_knowledge ($11.06\%$) methods. A Wilcoxon signed-rank test (The detailed results can be seen in \href{https://doi.org/10.5281/zenodo.14890689}{Digital Appendix}) confirmed that these improvements were statistically significant ($p < 0.001***$) for all methods except RAG\_CoT\_knowledge, which showed only a modest, non-significant gain of 3.36\% ($p = 0.07$). Notably, in the second-round feedback, RAG\_CoT\_learner achieved the highest correct evaluation rate $97.6\%$, reflecting an $6.73\%$ improvement ($p < 0.001***$).
\begin{table}
\centering
\scriptsize
\caption{Comparison of evaluation accuracy for feedback generated in the 1st and 2nd rounds}
\resizebox{1\textwidth}{!}{%
\renewcommand{\arraystretch}{1.25}
\label{tab5}
\begin{tabular}{l|c|c}
\hline
\textbf{Methods} & \textbf{Accuracy of 2nd feedback}& \textbf{Increase in accuracy compared to 1st feedback}\\
\hline
Baseline none & 96.63\% & +7.69\% \\
Baseline learner & 95.19\% & \textbf{+12.98\%} \\
Baseline knowledge & 94.71\% & +11.06\%\\
Rag cot none & 97.12\% & +2.89\% \\
Rag cot learner & \textbf{97.60\%} & +6.73\% \\
Rag cot knowledge & 96.63\% & +3.36\% \\
\hline
\end{tabular}
}
\end{table}

\begin{figure}
\includegraphics[width=\textwidth, trim=0cm 0.7cm 0cm 0cm, clip]{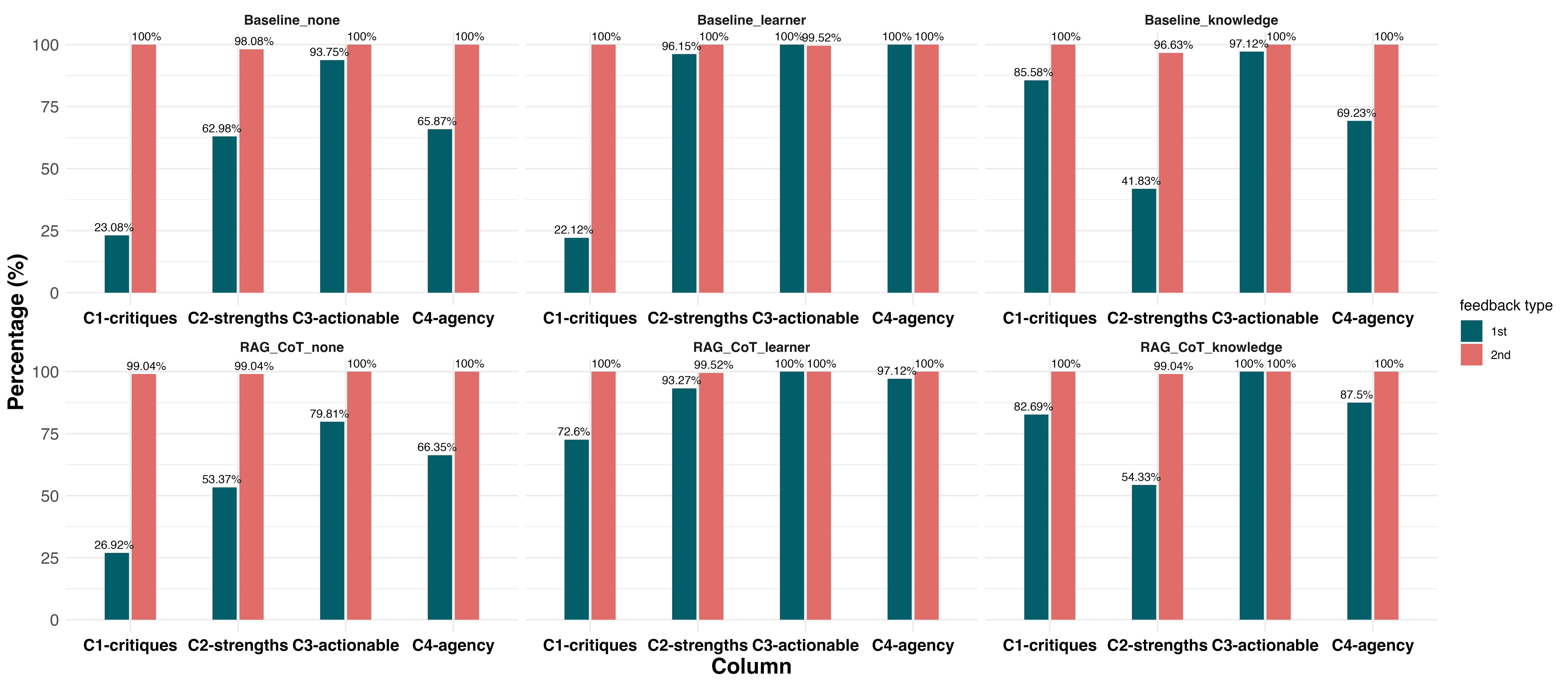}
\caption{Comparison of the percentage of components in feedback generated in the 1st and 2nd rounds} \label{fig4}
\end{figure}

In terms of the comparison of \textbf{components }dimension, Fig. \ref{fig4} demonstrates that the second-round feedback generated by six methods achieved high coverage across the four components, ranging from 99.04\% to 100\%. In addition, the percentage of second-round feedback generated by six methods containing all four components simultaneously ranges from 96.63\% to 99.52\%  (see detailed results in the \href{https://doi.org/10.5281/zenodo.14890689}{Digital Appendix}).  The Wilcoxon signed-rank test showed that feedback regenerated in the second round statistically significantly improved the deficiencies in C1-critiques ($p<0.001***$), C2-strengths($p<0.001***$ ; $p=0.005**$), C3-actionable ($p<0.001***$ ; $p=0.014*$), and C4-agency ($p<0.001$ ; $p=0.014*$) observed in the first-round feedback among different methods (see detailed test results in the \href{https://doi.org/10.5281/zenodo.14890689}{Digital Appendix}). Except for the learner framework condition, which already performed well in C3-actionable ( baseline\_learner: $p=0.317>0.05$; RAG\_CoT\_learner: $p=NA$) and C4-agency ( Baseline\_learner:$p=NA$), resulting in no improvements. In addition, under knowledge framework condition, RAG\_CoT\_knowledge also already performed well in C3-actionable ($p=NA$), showing in no improvements. This means that, after performing the ``G-E-RG'' approach, regardless of the prompt technique or whether a framework was applied in the first stage, it effectively enhances feedback, bringing all feedback to a balanced state and ensuring the presence of all four effective components. 

\begin{figure}
\includegraphics[width=\textwidth, trim=0cm 0cm 0cm 1cm, clip]{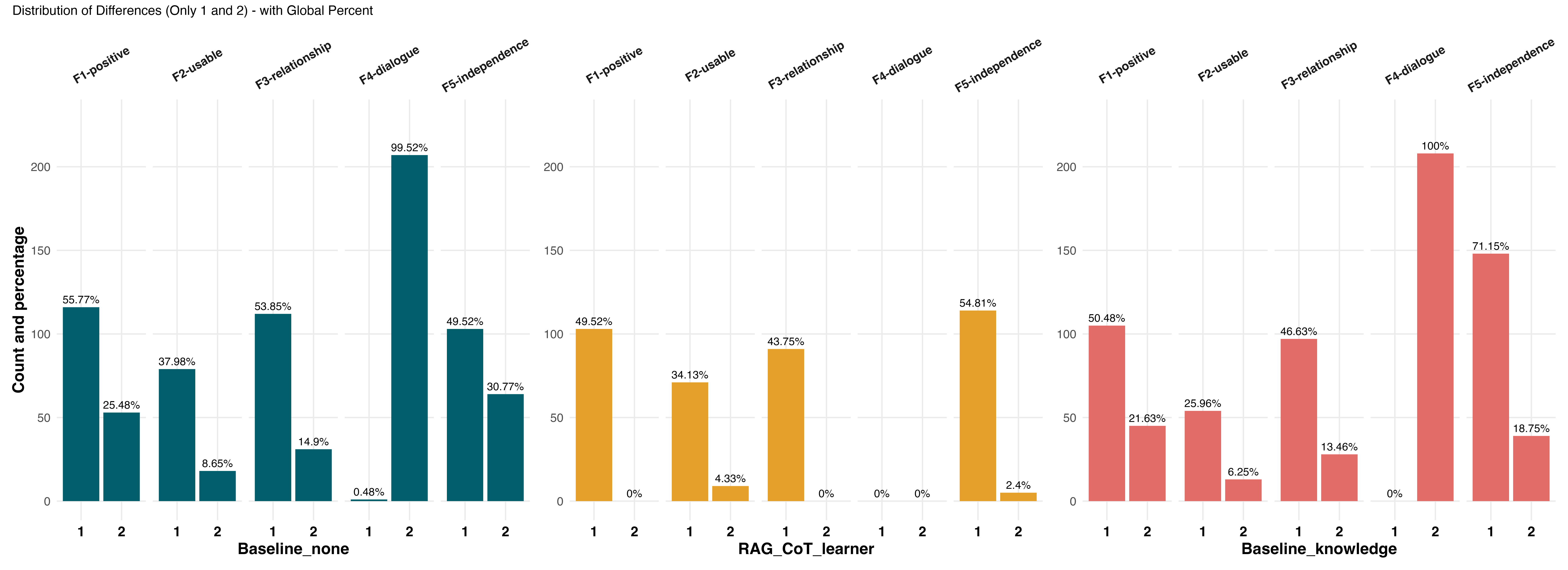}
\caption{Count and percentage distribution of the value increases (by 1/2) for five features in the 2nd-round feedback under three methods} \label{fig5}
\end{figure}

Regarding the \textbf{features }dimensions, the Wilcoxon signed-rank test results indicated that—with the exception of the Baseline\_learner ($p=0.317$) and RAG\_CoT\_learner ($p=NA$) method, which already performed exceptionally well in the F4-dialogue sub-dimension —all other methods and features showed significant improvements ($p<0.001***$) (See detailed test results in the \href{https://doi.org/10.5281/zenodo.14890689}{Digital Appendix}). As shown in Fig. \ref{fig5}, for the RAG\_CoT\_learner method—which initially exhibited relatively high feedback quality—the ``G-E-RG'' approach resulted in more modest improvements. For example, $38.46\% $of the feedback showed improvement in F2-usable (with $34.13\% $increasing by $1$ and the remaining by $2$), resulting in a mean score increase from $2.21$ to $2.53$. Similarly, approximately $49.52\% $of the feedback improved in F1-positive (all increasing by 1), leading to a mean score rise from $2.07$ to $2.54$. No improvement was observed in the F4-dialogue, as the initial mean score was already at 3. In contrast, the Baseline\_none and Baseline\_knowledge methods initially yielded relatively poor performance across all five features. After applying ``G-E-RG'', significant improvements were observed in all features, particularly in F4-dialogue, where nearly $100\%$ of the feedback reached the highest enhancement. However, it is worth noting that regardless of the method, the improved feedback generated in the 2nd-round still has room for enhancement in F1-positive, F2-usable, and F3-relationship, with their feature means ranging from $2.39/3$ to $2.63/3$. Specific examples of these feedback samples are also provided in the \href{https://doi.org/10.5281/zenodo.14890689}{Digital Appendix}.
Table \ref{tab6} presents a Pair-t test result of the \textbf{simplicity }dimension based on the word count of feedback. The ``Mean change'' column represents the change in the average word count of feedback in the second round relative to the feedback in the first round. With the exception of the Baseline\_knowledge ($p=0.439$) methods, all methods showed significant changes in word count. Specifically, while the RAG\_CoT\_none method experienced a significant increase in $20.34$ words per feedback ($p<0.001***$), the RAG\_CoT\_knowledge, Baseline\_learner, and RAG\_CoT\_learner methods exhibited significant decreases—reducing the average word count by $76.83$ ($p<0.001***$), $25.77$ ($p<0.001***$), and $20.34$ ($p<0.001***$) words per feedback instance, respectively. These results indicate that, combining other dimensions,  through evaluation and regeneration after the initial generation, longer feedback can be effectively condensed into a more concise form without sacrificing the essential components and features.
\begin{table}[ht]
\centering
\scriptsize
\caption{Pair-t-test of word count of feedback generated in the 1st and 2nd rounds}
\label{tab6}
\renewcommand{\arraystretch}{1.25}
\begin{tabular}{l|c|c|c|c|c|c}
\hline
\textbf{Methods} & \textbf{Mean} & \textbf{SD}& \textbf{Mean change}& \textbf{t} & \textbf{p} & \textbf{r} \\
\hline
Baseline-none        & 98.05& 5.00& 2.52& -2.58 & \textbf{.011}  & -.179 \\
Baseline-learner     & 97.60& 3.86& -25.77& 25.01 & \textbf{$<$ .001} & 1.734 \\
Baseline-knowledge   & 98.48& 5.20& -.74& .78   & .439  & .054  \\
RAG\_CoT none        & 98.35& 5.55& 20.34& -12.46& \textbf{$<$ .001} & -.864 \\
RAG\_CoT learner     & 97.83& 4.29& -15.24& 13.50 & \textbf{$<$ .001} & .936  \\
RAG\_CoT knowledge   & 97.05& 9.51& -76.83& 23.65 & \textbf{$<$ .001} & 1.640 \\
\hline
\end{tabular}
\end{table}

\section{Discussion and Conclusion}
This study introduced the ``G-E-RG'' method—a novel multi-agent process involving generation, evaluation, and regeneration—to enhance the quality of feedback produced by LLMs. Our findings demonstrated that the G-E-RG process significantly improved multiple dimensions (e.g., evaluation accuracy, C2-strengths, C3-actionable, F4-dialogue) of feedback quality regardless of applied methods, although the final quality of feedback in the 2nd round in F1-positive, F2-usable, and F3-relationship sub-dimensions remained suboptimal. In other words, regardless of the quality of the initial feedback, through evaluation and regeneration, it can be transformed into high-quality feedback in the second round,  which reduces the instability of LLM-generated content, ensures quality control of feedback, and enhances its applicability in real-world educational settings. This suggests that the ``G-E-RG'' method is extremely effective in aligning feedback with higher quality. However, the multi-agent pipeline is resource-intensive, which may increase the cost. In addition, further refinements or human involvement are necessary to achieve consistent high-quality performance across all dimensions, and exploring how to re-retrieve related slides based on evaluation results also may make a difference. Unlike prior work, which predominantly relied on one-round feedback generation (e.g., \cite{dai2024assessing,holderried2024language,wang2024chatgpt}). Our approach implemented an iterative process that refines feedback based on evaluation results. Although some studies have also investigated encoding feedback with LLMs (e.g., \cite{koutcheme2024open,phung2023generating}), they did not complete a cycle of re-generation informed by systematic evaluation. In this way, our study lays a new empirical foundation for integrating multi-agent iterative processes into automated feedback systems, enabling continuous improvement, dynamic quality monitoring, and learner-centered optimization of LLM-generated responses. 

An additional contribution of this study is the comparative analysis of prompt strategies in the first round of feedback generation. The RAG\_CoT method, which integrated external expert knowledge through retrieval-augmented generation and employs step-by-step reasoning, outperformed the baseline in the first-round feedback generation —aligning with recent findings (e.g., \cite{fung2024automatic,wei2022chain}). Additionally, our results revealed that LLM-generated feedback tends to be overly formulaic, often presuming student responses were incomplete regardless of responses' accuracy. Future research should explore methods that generate more varied, non-formulaic feedback and better differentiate between complete and incomplete responses. At the framework level, feedback generated using a learner-centered approach exhibits more comprehensive quality in the first round. This advantage likely stems from our evaluation rubric being rooted in learner-centered principles, which evaluate not only whether the information is effectively conveyed but also whether the feedback supports student autonomy and fosters meaningful teacher-student interaction -an essential aspect of learner-centered pedagogy. Embedding this framework thus ensures alignment with the learner-centered approach advocated by  \cite{goktas2024utilizing} and \cite{hwang2023exploring}.   However, there is a limitation in the current pipeline; once mismatched slides are retrieved, they continue to propagate errors in subsequent feedback generation because the same slides are reused. Further study should explore how to enhance the accuracy of retrieval.  In addition, the feedback accuracy is not 100\%, which means that further study still needs to explore more advanced mechanisms to bridge that gap. Our study revealed limitations in automated feedback evaluation. Although our system efficiently measured several dimensions of feedback quality, its accuracy was less reliable when assessing critical features such as F2-usable. Future research should investigate a human-in-the-loop framework in which humans can provide final assessments for flag instances or dimensions where LLMs under-perform to ensure the accuracy of feedback evaluation.

\noindent\textbf{Acknowledgments.} This research was supported by the Generative AI + Education Tools R\&D Seed Grant at Carnegie Mellon University.


%
%
%

\bibliographystyle{splncs04}
\bibliography{Ref}
\end{document}